\documentclass[preprint]{aastex}
\usepackage{bm}

\newcommand{\sub}[1]{_{\mbox{\scriptsize #1}}}

\shorttitle{Cross Sections of Dust Aggregates and a Compression Model}
\shortauthors{Toru Suyama}

\begin{document}
\title{Geometrical Cross Sections of Dust Aggregates and a Compression
Model for Aggregate Collisions}

\author{Toru Suyama\altaffilmark{1,2}, Koji Wada\altaffilmark{3},
Hidekazu Tanaka\altaffilmark{2}, and Satoshi Okuzumi\altaffilmark{4}}
\altaffiltext{1}{Nagano City Museum, Hachimanpara Historical Park
Ojimada-machi, Nagano 381-2212, Japan; museum@city.nagano.lg.jp}

\altaffiltext{2}{Institute of Low Temperature Science, Hokkaido
University, N19-W8, Sapporo 060-0819, Japan}

\altaffiltext{3}{Planetary Exploration Research Center, Chiba Institute
of Technology, Tsudanuma 2-17-1, Narashino, Chiba 275-0016, Japan}

\altaffiltext{4}{Department of Physics, Nagoya University, Nagoya, 
Aichi 464-8602, Japan}

%
\begin{abstract}
Geometrical cross sections of dust aggregates determine their coupling
with disk gas, which governs their motions in protoplanetary
disks. Collisional outcomes also depend on geometrical cross sections of
initial aggregates. In the previous paper, we performed
three-dimensional $N$-body simulations of sequential collisions of
aggregates composed of a number of sub-micron-sized icy particles and
examined radii of gyration (and bulk densities) of the obtained
aggregates.
We showed that collisional compression of aggregates is not 
efficient and that aggregates remain fluffy.
In the present study, we examine geometrical cross sections of the
aggregates. Their cross sections decreases due to the compression
as well as their gyration radii.
It is found that a relation between the cross section and the 
gyration radius proposed by Okuzumi et al. is valid for the compressed
aggregates.
We also refine the compression model proposed in our previous paper.
The refined model enables us to calculate the evolution of 
both gyration radii and cross sections of growing aggregates
and reproduces well our numerical results of sequential aggregate
collisions.
The refined model can describe non-equal-mass collisions as well as
equal-mass case.
Although we do not take into account oblique collisions in the present study, 
oblique collisions would further hinder compression of aggregates. 
\end{abstract}
\section{INTRODUCTION}

In the theory of planet formation, planets are thought to have formed in
protoplanetary disks through mutual collisions and coalescence of
planetesimals. The formation process of planetesimals, on the other
hand, still has a large uncertainty. Before the planetesimal formation,
dust grains grow through their collisional coalescence in a
protoplanetary disk and settle to the disk mid-plane,  which forms a
dense dust layer at the mid-plane \citep[e.g.,][]{saf69,nak81,tan05,dul05}.
Planetesimals would be formed in the dust layer through gravitational
instability \citep[e.g.,][]{gol73,sek98,you02},
streaming instability \citep[e.g.,][]{you05,you07}
or simple coalescence \citep[e.g.,][]{wei93,bra08a,bra08b}.
In these models of planetesimal formation, motion of dust
grains is an important factor because it determines the spatial
distribution of dust grains and the collision speed between
them. Furthermore, their motion is governed by the drag forces from the
disk gas.

Gas drag forces on dust grains strongly depend 
on their internal
structure (or their bulk densities). Most studies of dust growth in
protoplanetary disks have assumed compact structure of dust
grains. However, dust grains 
growing
through mutual
collisions would actually be aggregates of (sub-micron) primitive grains
and the aggregates have a fluffy structure with an extremely low bulk
density, as reported by experimental and theoretical studies
\citep[e.g.,][]{blu04b,orm07,suy08,oku09,zsom11}.
Such fluffy aggregates
have large ratios of their 
geometrical 
cross sections to masses, which
significantly enhance 
gas drag forces
on them compared with compact
dust grains. 
Hence, in order to clarify dust growth and planetesimal formation 
in protoplanetary disks, we have to examine the internal structure and 
the geometrical 
cross sections of dust aggregates 
(we use a term `cross section' in referring to `geometrical cross
section', hereafter). 

\citet{suy08} (hereafter S08) performed $N$-body numerical simulations of
sequential aggregate collisions to examine the compression process of
growing aggregates. 
The sequential collisions mean that we repeat collisions of
aggregates obtained at the previous collisions.
With such a simulation, we can observe a natural evolution of 
the aggregate structure.
Their numerical results showed that large aggregates have an
extremely low bulk density 
in spite of
compression at aggregate collisions. In the early stage of dust growth,
aggregates just stick without any restructuring because of their low 
impact energy and they have a fluffy structure with an extremely 
low bulk density as they grow. 
In the later stage in which the impact energy exceeds a
critical energy, aggregates are gradually compressed. 
Even in this compression stage, their density remains very low.
It is found that the compressed aggregates have a low fractal
dimension of 2.5. This structural feature causes the low density
of the compressed aggregates.
S08 also
derived a formula describing the density evolution of growing
aggregates. To estimate their bulk densities, S08 
used
the
so-called gyration radii of the aggregates but did not examine their
cross sections. 
However, cross sections are directly related to the gas drag 
forces
rather than gyration radii. 
It is necessary to clarify the evolution of
cross sections of dust aggregates during their growth. 

\begin{figure}
 \plotone{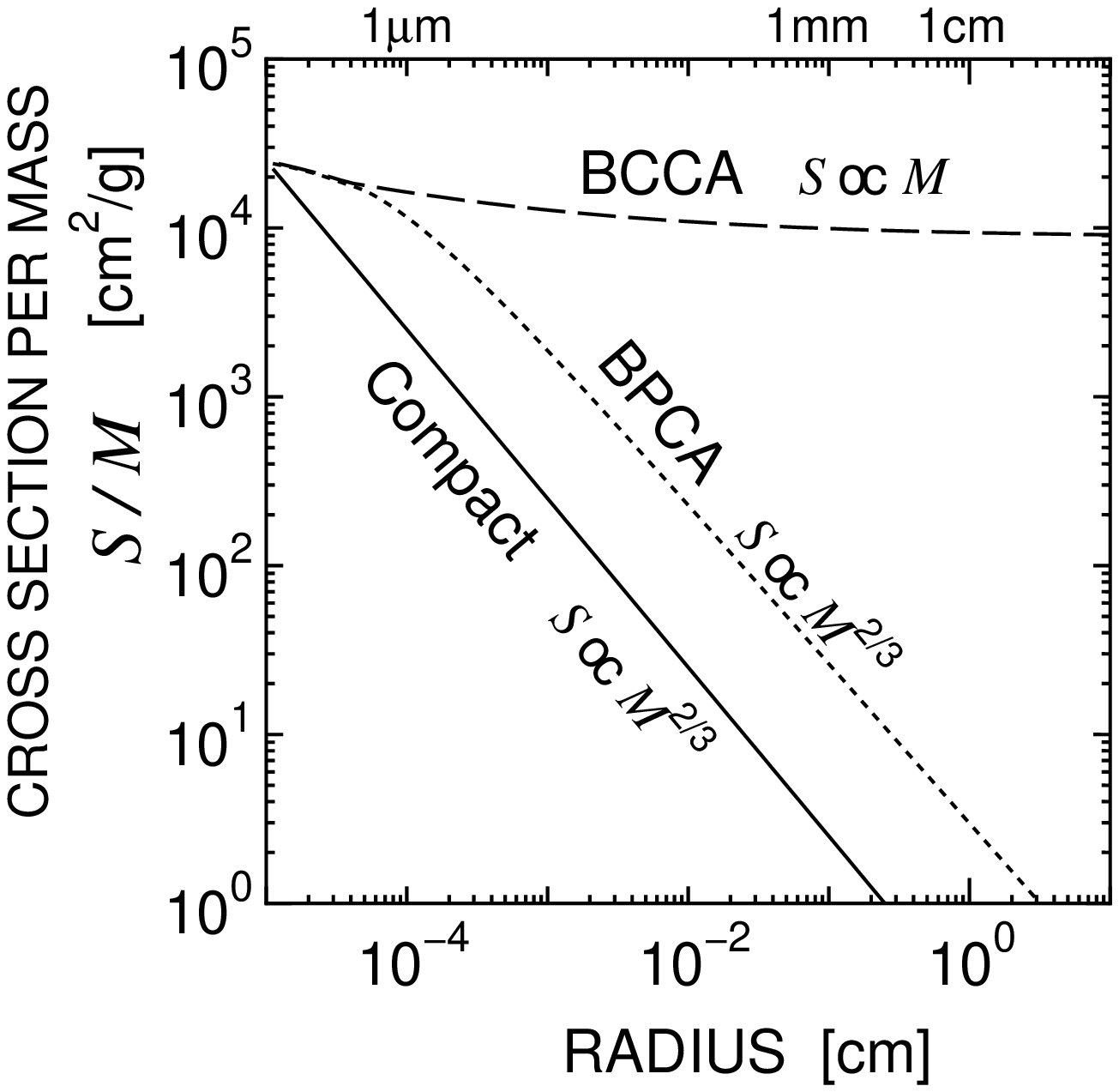}
 \caption{
 Geometrical cross sections per masses for various types of aggregates.
 Constituent monomers are 0.1$\mu$m-seized icy spherical particles.
 The solid line shows the geometrical cross section of
 a compact sphere with the same density as constituent particles.
 The dotted line and the dashed line show the geometrical cross sections
 of BPCA and BCCA clusters, respectively.
 For BCCA and BPCA, we used empirical formulae by \cite{min06b}.
 }
 \label{fig:spmintro}
\end{figure}

Cross sections of aggregates depend on their internal
structure. There are two simple aggregate models. 
One is Ballistic Cluster-Cluster Aggregation (BCCA). 
A BCCA cluster is formed through
collisions between 
two equal-sized clusters. 
Second is Ballistic Particle-Cluster Aggregation (BPCA). 
A BPCA cluster is formed through
deposition of small monomer particles on a large cluster. 
For both BCCA and BPCA, restructuring is assumed to be
negligible at each collision.
The BCCA clusters have very fluffy and open structures and 
the BPCA clusters have relatively
compact structures. 
Figure~\ref{fig:spmintro} shows the ratios of cross
sections to masses of aggregates. 
It is shown that the cross section per mass strongly
depend on aggregate types. 
Cross sections of dust aggregates are expected to be between those of
BCCA and BPCA clusters. 
One may consider that a cross section is approximately given by the
square of a gyration radius.
Cross sections of aggregates are, however, generally
independent of their gyration radii, especially for highly fluffy
aggregates.
The non-dimensional ratio of the cross section to the square
of the gyration radius gradually decreases with their growth for BCCA
clusters (Minato et al.~2006; see also Fig.~6).
On the other hand, this ratio is almost constant in the growth
of BPCA clusters. 
Okuzumi et al. (2009) proposed a useful relation between 
the cross section and the gyration radius
for various aggregates formed through 
hit-and-stick growth (as well as BCCA and BPCA). 
Paszun and Dominik (2009) also derived another relation.
Nevertheless, it is not clear whether these relations are
also valid for aggregates compressed at collisions.
We check the validity of these relations,
using the resultant aggregates obtained by S08.

Once we find a valid relation between the cross sections and the 
gyration radii, it would be very helpful to describe the evolution 
of the cross sections because the compression model by S08
can describe gyration radii of growing aggregates.
The compression model by S08, however, has some limitations.
This model is not directly applicable to low-energy collisions
(i.e., hit-and-stick collisions) or to non-equal-mass collisions.
In order to describe gyration radii and cross sections of aggregates
for all growth stages seamlessly, we further refine the compression model, 
by removing these limitations.

%

In laboratory experiments, \citet{weidl09}
examined compression of aggregates consisting of 
1.5$\mu$m-diameter SiO$_2$ spheres at their multiple 
rebounds and also developed an empirical compression model. 
Initial aggregates in their experiments
possess a volume filling factor of $\sim 0.1$, 
which is approximately equal to 
that of BPCA clusters. On the other hand, S08 and the 
present paper focus on the compression of fluffier 
aggregates of which filling factor is between BCCA and 
BPCA clusters during their collisional growth. 
Hence our compression model and theirs are
complementary to each other. As mentioned above, dust 
aggregates are expected to have much smaller bulk densities
than BPCA clusters at the early stage of their 
growth in protoplanetary disks. Our compression model is 
useful as long as bulk densities of aggregates is lower 
than that of BPCA clusters.

It should be noticed that S08 only considered head-on collisions of
aggregates in their numerical simulations of sequential
collisions. The oblique collisions are expected to hinder the
compression (Wada et al. 2007; Paszun and Dominik 2009).
In the present study, however, we use the results in S08 as the 
first step. We will examine the effects of oblique collisions 
in future work.

In the next section, we briefly summarize the results of S08. In Section
3, we numerically 
calculate
cross sections for the aggregates obtained by S08. 
We find that Okuzumi et al's relation between
the cross section and the gyration radius is valid
for compressed aggregates, too.
In Section 4, we refine the compression model by S08,
by removing its limitations in a reasonable way.
We find that the refined compression model reproduces well 
both of gyration radii and cross sections of aggregates
obtained by the numerical simulation,
with the help of Okuzumi et al's relation.
We also check the validity of the refined model for non-equal-mass 
collisions with additional numerical simulations of aggregate collisions.
A summary is given in the last section.
\section{RESULTS OF AGGREGATE COMPRESSION IN $N$-BODY SIMULATIONS BY S08}

Suyama et al.~(2008) performed $N$-body numerical simulations of head-on
aggregate collisions and examined the density evolution of aggregates
growing through the collisions. We examine the cross section
of the resultant aggregates obtained by S08. Before
that, we briefly describe the numerical results of S08.

In the simulations, aggregates consist of a large number of icy spherical
particles with a radius of $r_1=0.1\mu$m.
S08 adopted the particle-interaction
model by Wada et al.~(2007). In the interaction model, 
repulsive and adhesive forces in the normal direction between particles
in contact
are given by the JKR theory (Johnson et al. 1971).
A
tangential force and a torque also arise to
resist the slide, roll, and twist motions between them. Aggregate
compression is regulated mainly by inelastic rolling motions of the constituent
particles (e.g., Dominik \& Tielens 1997;
Wada et al. 2007, 2008 [hereafter W07,W08]~; G\"{u}ttler et al. 2010).
The rolling energy $E_\mathrm{roll}$
is the energy required for rolling 
a particle on its contact neighbor by an angle of $\pi/2$.
The rolling
energy is given by (W07, S08)
\begin{equation}
 E_\mathrm{roll}=6\pi^2\gamma r_1\xi_\mathrm{crit},\label{eroll} 
\end{equation}
where $\xi_\mathrm{crit}$ is the critical displacement for inelastic
rolling motion.
The parameter range of $\xi_\mathrm{crit}$ is set to be from 2 to 16
$\mathrm{\mbox{\AA}}$ in S08. 
A large
rolling energy $E_\mathrm{roll}$
suppresses the restructuring of
aggregates. 
To examine the structure evolution of
growing aggregates, S08 performed $N$-body simulations of sequential
collisions. Each
simulation starts from a collision of aggregates composed of two
particles (i.e., dimers) and ends with a collision of aggregates
composed of 16,384 particles. 
The resultant aggregate obtained in the previous
collision is used as initial aggregates at each collision in the
simulation of sequential collisions.
The impact velocity is constant in sequential collisions. 
For various (constant) impact velocities and critical rolling displacements,
they performed a large number of runs of sequential collisions.

As an index of the 
size of an aggregate,
S08 adopted the
radius of gyration, $r_g$, defined by
\begin{equation}
 r_g\equiv
  \sqrt{\sum_{i=1}^N
  \frac{|\mbox{\boldmath{$x$}}_{i}-\mbox{\boldmath{$x$}}_\mathrm{M}|^2}
  {N}
  },
\end{equation}
where $\mbox{\boldmath{$x$}}_i$ is the position of particle $i$,
$\mbox{\boldmath{$x$}}_\mathrm{M}$ is
the position of the center of mass of the aggregate, and $N$ is the
number of particles composing the aggregate.
Using the radius of gyration, 
the volume $V$ and 
the bulk density 
$\rho$ 
of the aggregate are 
evaluated to be (Mukai et al. 1992; W08)
\begin{equation}
 V(r_g)=\frac{4\pi}{3}\left(\sqrt{\frac{5}{3}}r_g\right)^3,
\label{vdef}
\end{equation}
\begin{equation}
 \rho(r_g)=\frac{m_1N}{V(r_g)},
\label{rhodef}
\end{equation}
respectively,
where 
$\sqrt{5/3}r_{g}$ is the so-called characteristic radius of an aggregate
and 
$m_1$ is the mass of a constituent particle.

Aggregates are expected to have a BCCA structure for collisions at
sufficiently low velocity because of sticking together of equal-mass
aggregates without any restructuring. If the compression is effective at
collisions, the gyration radii of aggregates would become smaller than
those of BCCA clusters. It is meaningful to compare the
obtained aggregates with BCCA clusters. 
Since BCCA clusters have a fractal dimension of $\sim 2$, the radius of
gyration of  the BCCA cluster is given for large $N$ by (e.g., Mukai et
al. 1992; W08
)
\footnote{Exactly speaking,
equation (\ref{bccarg}) is satisfied for BCCA clusters formed through
head-on (hit-and-stick) collisions, which have the fractal dimension
of 2.0. When offset collisions are also included at the formation
of BCCA, their fractal dimension is 1.9 and the gyration radii
proportional to $N^{0.52}$ (Okuzumi et al. 2009). 
Although the later BCCA is more realistic,
the former BCCA is used in S08 and the present study 
since S08 consider only 
head-on collisions in their simulations.}

\begin{equation}
 r_{g,\mathrm{BCCA}}\simeq N^{0.50}r_1.\label{bccarg}
\end{equation}

\begin{figure}
 \epsscale{0.6}
 \plotone{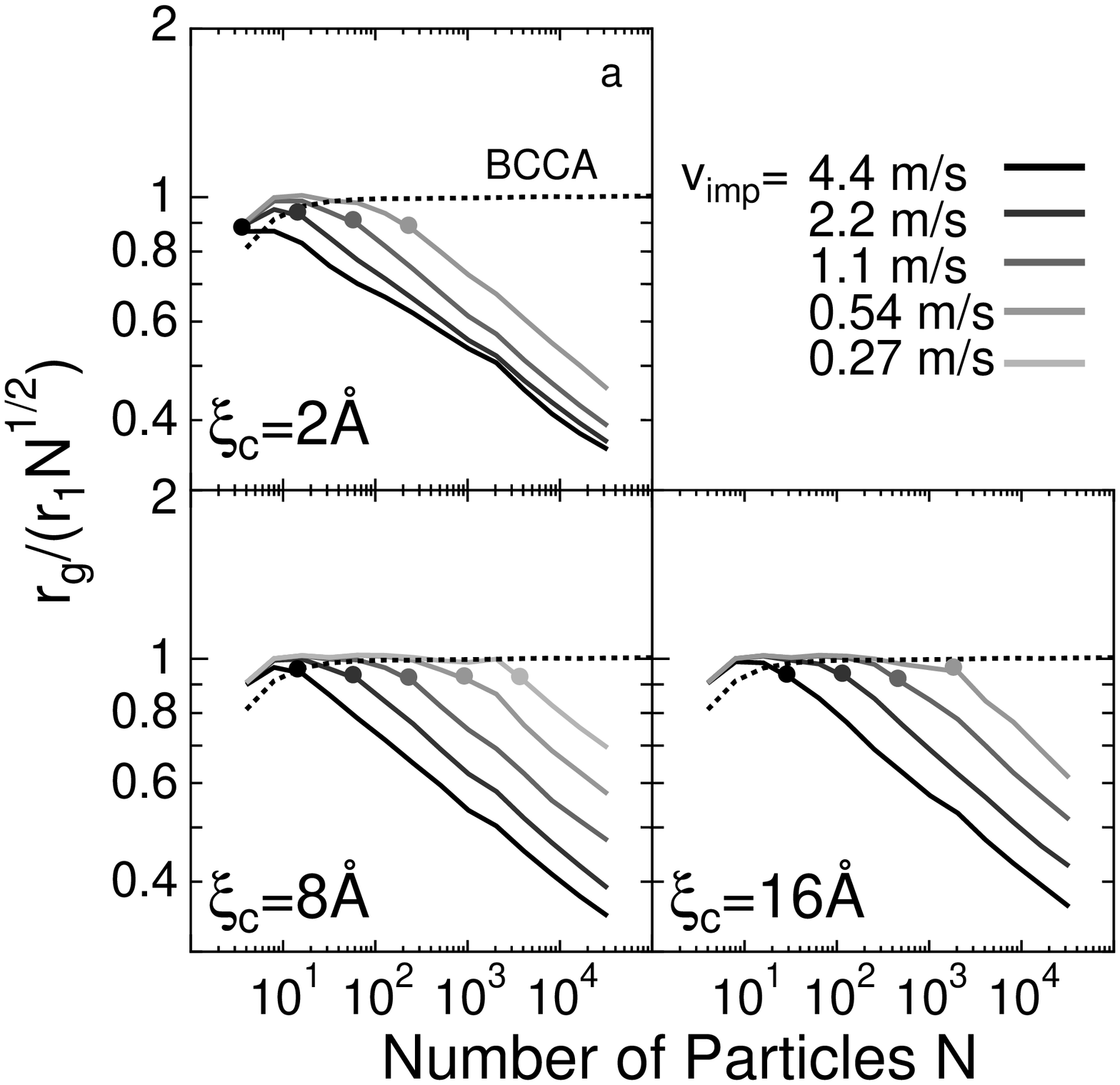}\\
 \plotone{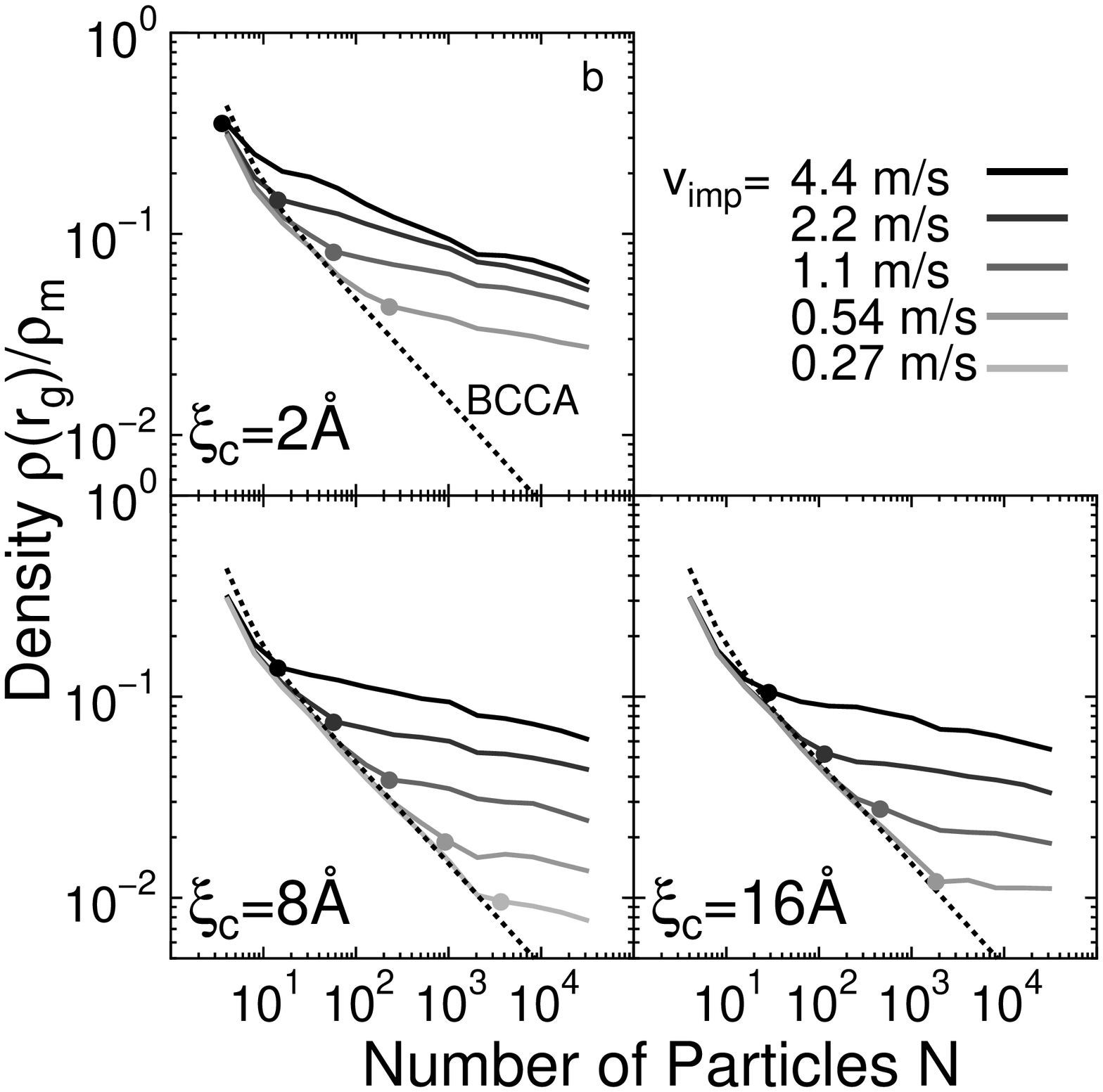}
 \caption{Structure evolution of growing aggregates in the sequential
 collision 
 simulations for various impact velocities $v_\mathrm{imp}$ and 
 critical rolling displacements $\xi_\mathrm{crit}$. 
 Panel~(a) shows the radius of gyration and panel~(b) shows the density,
 respectively. The solid lines show
 the resultant aggregates in our simulations, and the dashed
 lines indicate BCCA clusters. Filled circles indicate the
 critical number of particles $N_\mathrm{crit}$ to start
 compression, as estimated from equation~(\ref{ncrit}) with
 $\beta=0.5$.}
 \label{fig:rgrho}
\end{figure}

Figure~\ref{fig:rgrho}a shows the gyration radius of the aggregates in the
simulations of sequential collisions performed by S08 for various values of
parameters, $\xi_\mathrm{crit}$ and the impact velocity
$v_\mathrm{imp}$.
The density of monomer particles is given by 
$\rho_m (\equiv 3m_1/[4\pi r_1^3])$. 
In the simulation of sequential collisions, the size of growing
aggregates is dependent on 
the direction of each collision.
S08 did 30 runs of the simulation of sequential
collisions and obtained the averaged value of $r_g$ from 30 runs for each
$\xi_\mathrm{crit}$ and $v_\mathrm{imp}$.
In Figure~\ref{fig:rgrho}a,  the horizontal axis is the number of the
constituent particles, $N$, in the growing aggregates and the vertical
axis is the gyration radii divided by $N^{1/2}r_1$ for comparison
with BCCA clusters.
The dashed line represents the radius of the BCCA cluster
and it is 
almost flat for
large $N$
as expected from equation (\ref{bccarg}).
The size of small aggregates produced in our simulation is almost the same as
that of BCCA clusters. 
This is because
the impact energy is small enough at the early stage of the aggregate
growth and the compression is ineffective at
each collision. As the aggregates grow,
the impact energy increases.
When the impact energy attains to $E_\mathrm{roll}$, the compression of
the aggregate starts: aggregates become smaller than BCCA clusters. 
The critical number of
particles, $N_\mathrm{crit}$, in the aggregate for compression is given
by (W08, S08)
\begin{equation}
 N_\mathrm{crit} = \beta\frac{8E_\mathrm{roll}}{m_1v_\mathrm{imp}^2},
\label{ncrit}
\end{equation}
where $\beta$ is a non-dimensional coefficient. In Figure~\ref{fig:rgrho}a,
we also plot the critical number $N_\mathrm{crit}$ with filled circles on
each curve, by setting $\beta=0.5$.
Figure~\ref{fig:rgrho}b shows bulk densities of growing aggregates in
these simulations, which evaluated with $r_g$ by
equation~(\ref{rhodef}). 
We also plot the density of BCCA clusters.
It decreases as $N^{-0.50}$ for large $N$.
After the onset of compression (i.e., $N > N_\mathrm{crit}$), the
bulk densities of the aggregates obtained by the numerical simulation 
are larger than that of the BCCA but still keep on decreasing gradually
in all cases,
indicative of the inefficiency of collisional compression.
S08 also developed the compression model,
which reproduces 
the density evolution of growing aggregates in
the compression stage. We will describe the compression model 
in Section 4.
\section{GEOMETRICAL CROSS SECTIONS OF AGGREGATES PRODUCED IN THE
 SEQUENTIAL COLLISIONS}

\begin{figure}
 \epsscale{1.0}
 \plotone{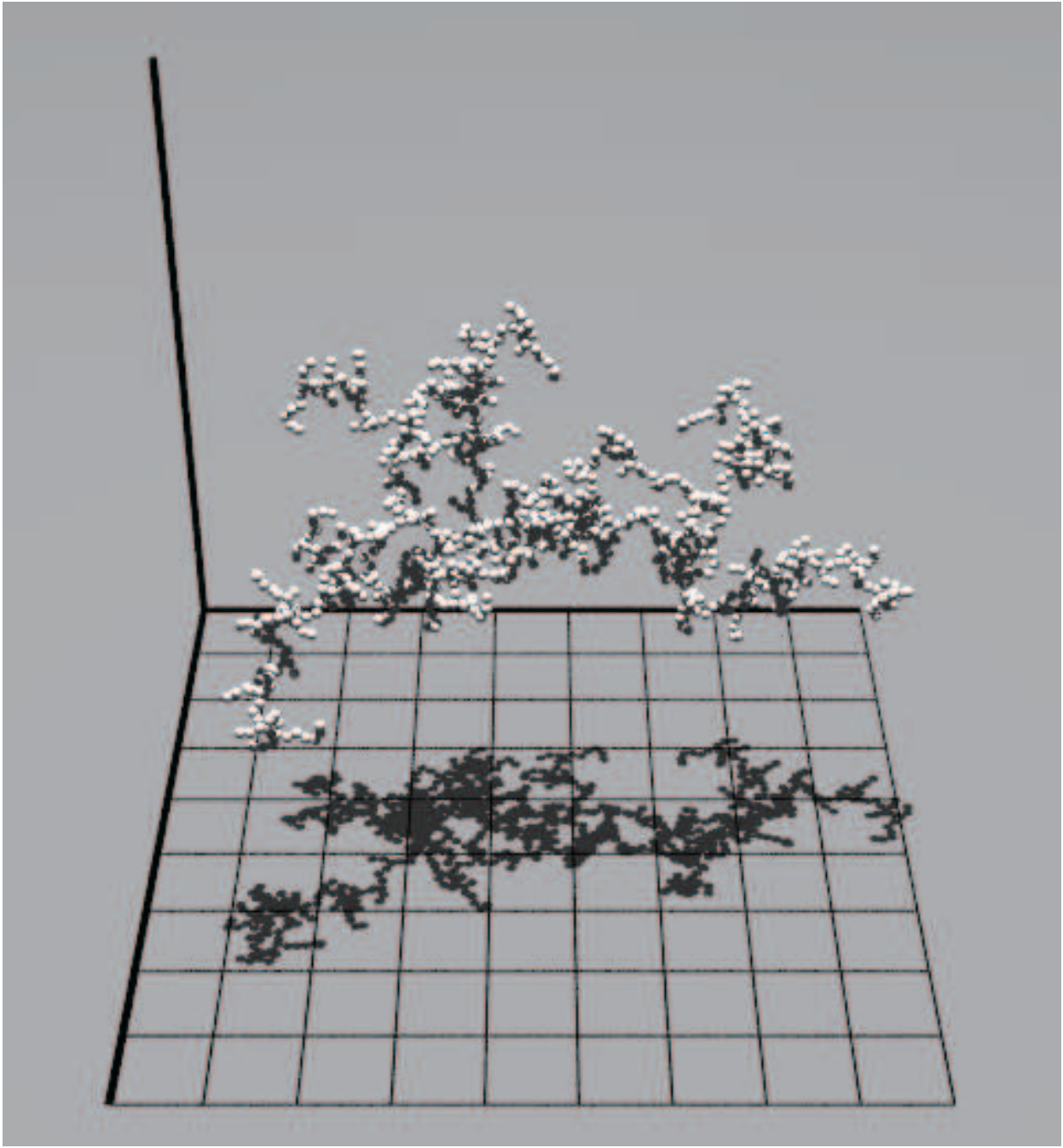}
 \caption{A schematic view of the method of the calculation of the
 geometrical cross
 section of dust aggregates. The shadow on a plane expresses the geometrical 
 cross section of the aggregate. 
 }
 \label{fig:ssmosikizu}
\end{figure}

\begin{figure}
 \epsscale{0.6} 
 \plotone{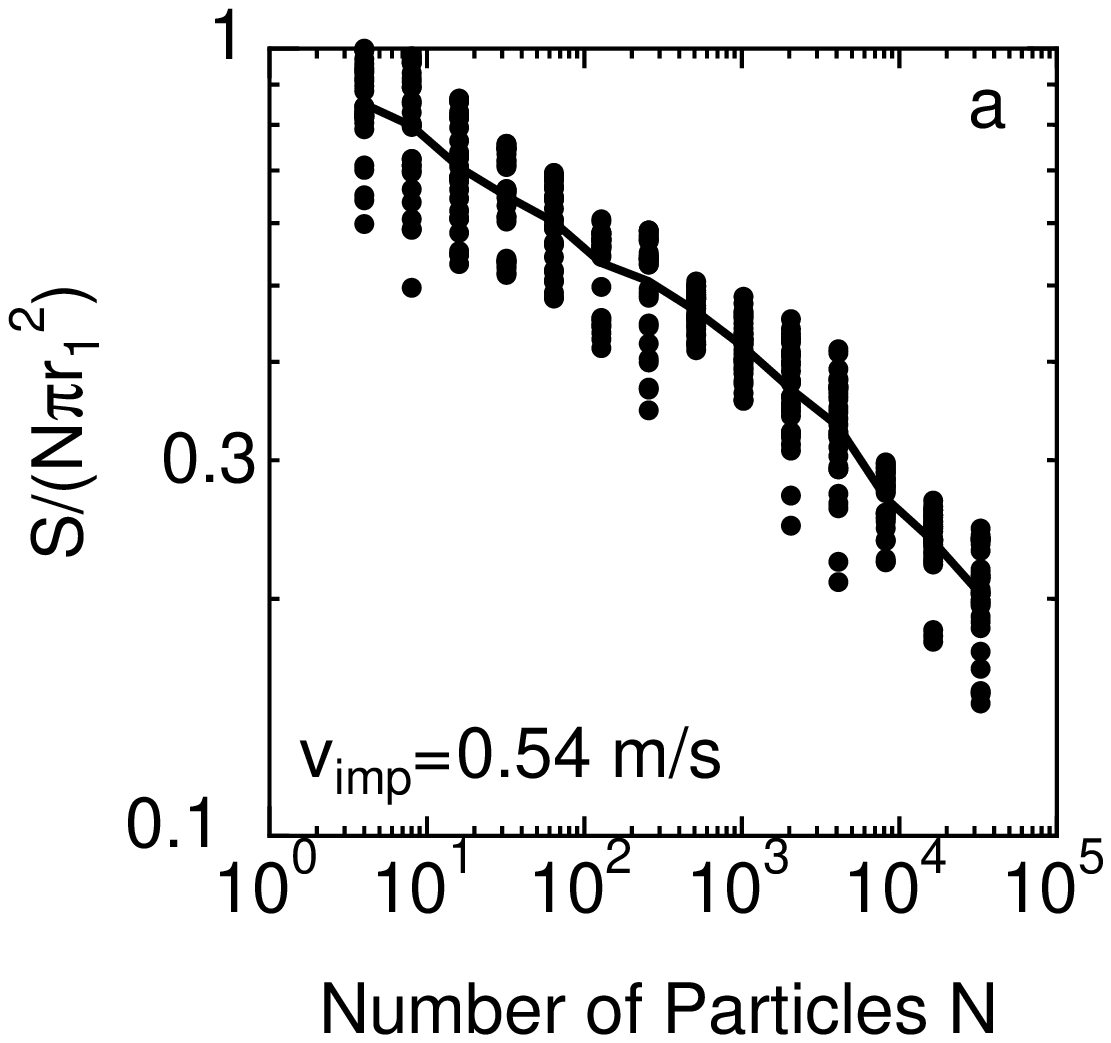}\\
 \plotone{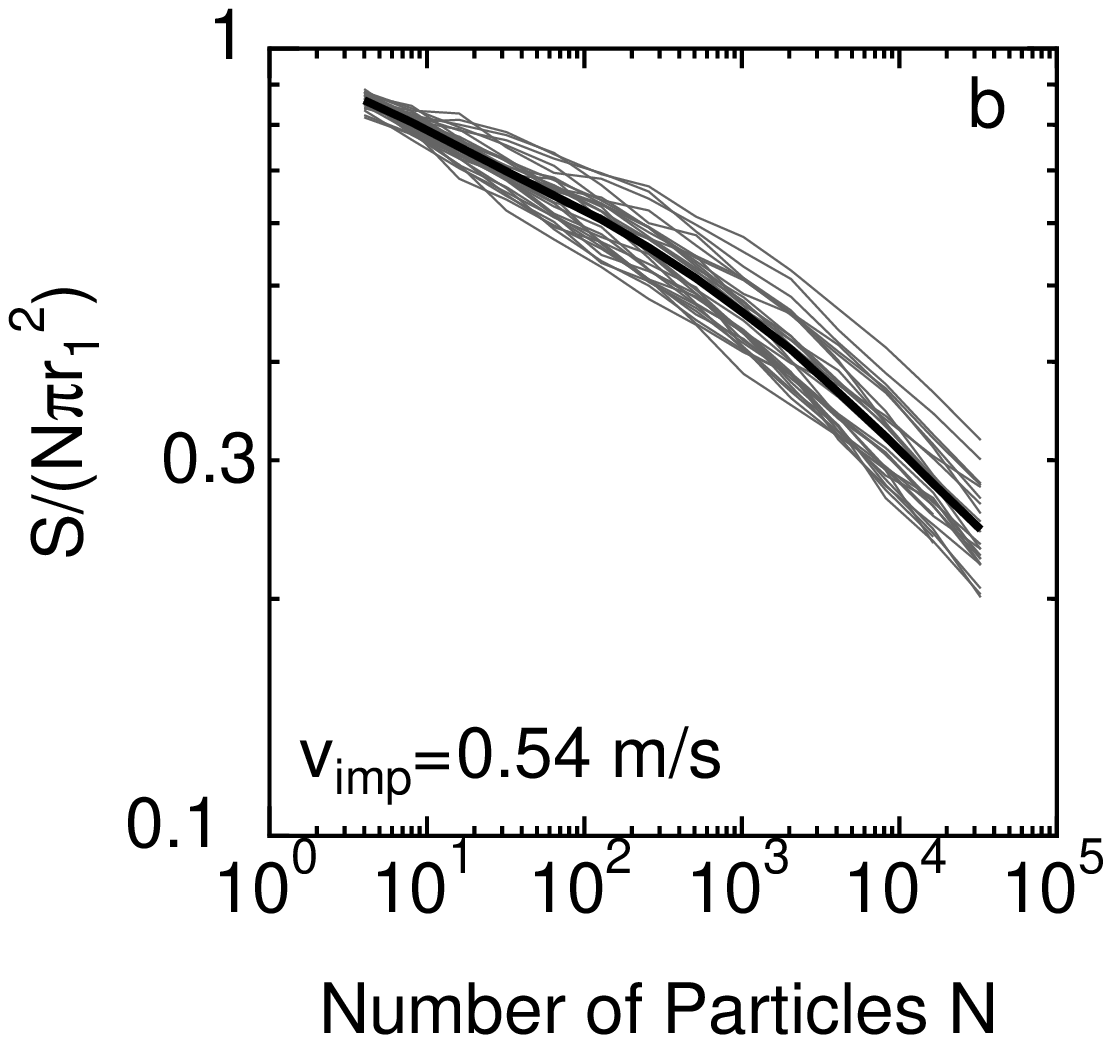} 
 \caption{ Geometrical cross
 sections of the resultant aggregates in the case of
 $v_\mathrm{imp}=0.54$ m~s$^{-1}$ and $\xi_\mathrm{c}=2
 \mathrm{\mbox{\AA}}$ as a function of the number of particles $N$. In
 panel (a), the dots show geometrical cross sections for 30 directions
 and the line shows the mean geometrical cross section. In Panel (b),
 thin gray lines show mean geometrical cross sections calculated for 30
 resultant aggregates and the solid line shows the average value of
 them. We obtain evolution of the geometrical cross section with these
 two kinds of averaging.  } \label{fig:ss1headon}
\end{figure}

We numerically 
calculate
cross sections of resultant aggregates 
obtained by
S08. The cross section of a dust aggregate is given by
the area of the shadow of the aggregate projected onto a plane. The area
of the shadow is calculated by counting the number of square meshes
in the shadow (Fig.~\ref{fig:ssmosikizu}). The width of the square
meshes is set to be
$0.0055 r_1$. This width is much smaller than the radius of the
monomer particle, $r_1$, 
though meshes with a much wider width are
drawn to emphasize them in Figure~\ref{fig:ssmosikizu}.
The area of the shadow is dependent on the
plane onto which the shadow is projected.
We calculate the areas of the shadows for 30 orientations randomly
chosen and define the cross section of the aggregate
by the mean values of the areas. Figure~\ref{fig:ss1headon} shows the
cross section
calculated in this way for  aggregates produced in the simulations of
sequential collisions. 
The vertical axis is the cross
section divided by $N \pi r_1^2$, 
which corresponds to the (non-dimensional) cross section per mass. 
If the overlapping of the monomer particles in the shadow is negligibly 
small, the value of the vertical axis 
approaches
unity. A filled circle in Figure~\ref{fig:ss1headon}a 
indicates the shadow area
for each orientation and the line shows the mean of them. 
The cross section per mass
decreases as an aggregate grows in the simulation of sequential
collisions due to the overlapping of the constituent particles. 
Since S08 did 30 independent runs of sequential collisions, the mean cross
sections are calculated and plotted as thin lines in
Figure~\ref{fig:ss1headon}b. 
Then we obtain the averaged value of the
30 mean cross sections as shown by the thick line in
Figure~\ref{fig:ss1headon}b.
The dispersion of the mean geometrical cross section can be evaluated
in Figure~\ref{fig:ss1headon}b.
The standard deviation of the mean geometrical
cross section is equal to or less than 11\% of its averaged value
during the aggregate growth.
In this way, we did two kinds of averaging to calculate the cross section
of growing aggregates in the simulation.
We show and discuss the cross sections of aggregates by using these
finally-obtained cross sections hereafter.

\begin{figure}
 \epsscale{1.0} 
 \plotone{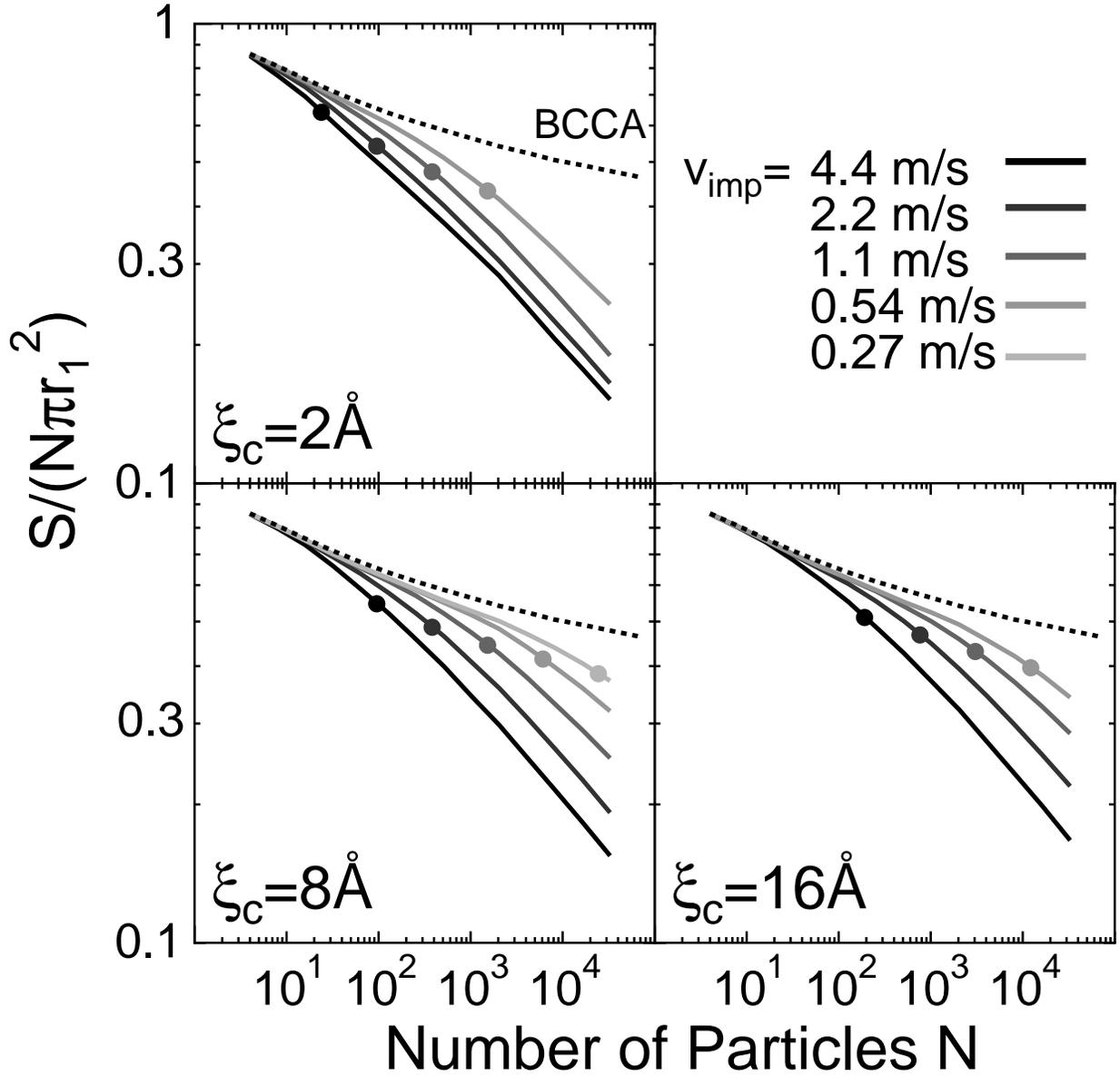}
 \caption{Geometrical cross sections per unit mass of growing aggregates
 in the simulations for various impact velocities $v_\mathrm{imp}$ and
 critical rolling displacements $\xi_\mathrm{c}$.
 The solid lines show the geometrical cross sections of the resultant
 aggregates in $N$-body simulations, and the dotted lines indicate the
 geometrical cross sections of BCCA clusters.  Collisional compression
 decreases their geometrical cross sections from the value of BCCA
 clusters. Filled circles indicate $N_\mathrm{crit}$ given by
 equation~(\ref{ncrit}) with $\beta=2.0$.  } \label{fig:ssheadon}
\end{figure}

Figure~\ref{fig:ssheadon} shows the cross section of resultant
aggregates for various values
of the parameter set ($\xi_\mathrm{crit}, v_\mathrm{imp}$).
The averaged cross section of the BCCA cluster is also calculated and
plotted. 
The cross section of BCCA we obtained agrees 
with the result of
\citet{min06b}.
Even for BCCA, the ratio $S/(N \pi r_1^2)$
gradually decreases with an increase in $N$ due to the overlapping of
constituent particles.
In the early stage of the aggregate growth (i.e. for small $N$), the cross
sections of the resultant aggregates change almost along the line of
BCCA. As the aggregates grow, however, their cross sections deviate from
the line of BCCA and become much smaller than that of BCCA, which
is due to compression at collisions. This qualitative tendency 
is consistent
with the evolution of the radius of gyration (Fig.~\ref{fig:rgrho}a).
Although the change in the cross sections is gradual and the starting
points of compression in the cross sections are not clear compared with
those in the gyration radii, the starting points are also described by
equation~(\ref{ncrit}), by setting the parameter
$\beta$ to be 2.0.
The larger value of $\beta$ than Figure~\ref{fig:rgrho} indicates that
the onset of compression in the cross section is later than that in the
gyration radius.

\begin{figure}
 \epsscale{1.0}
 \plotone{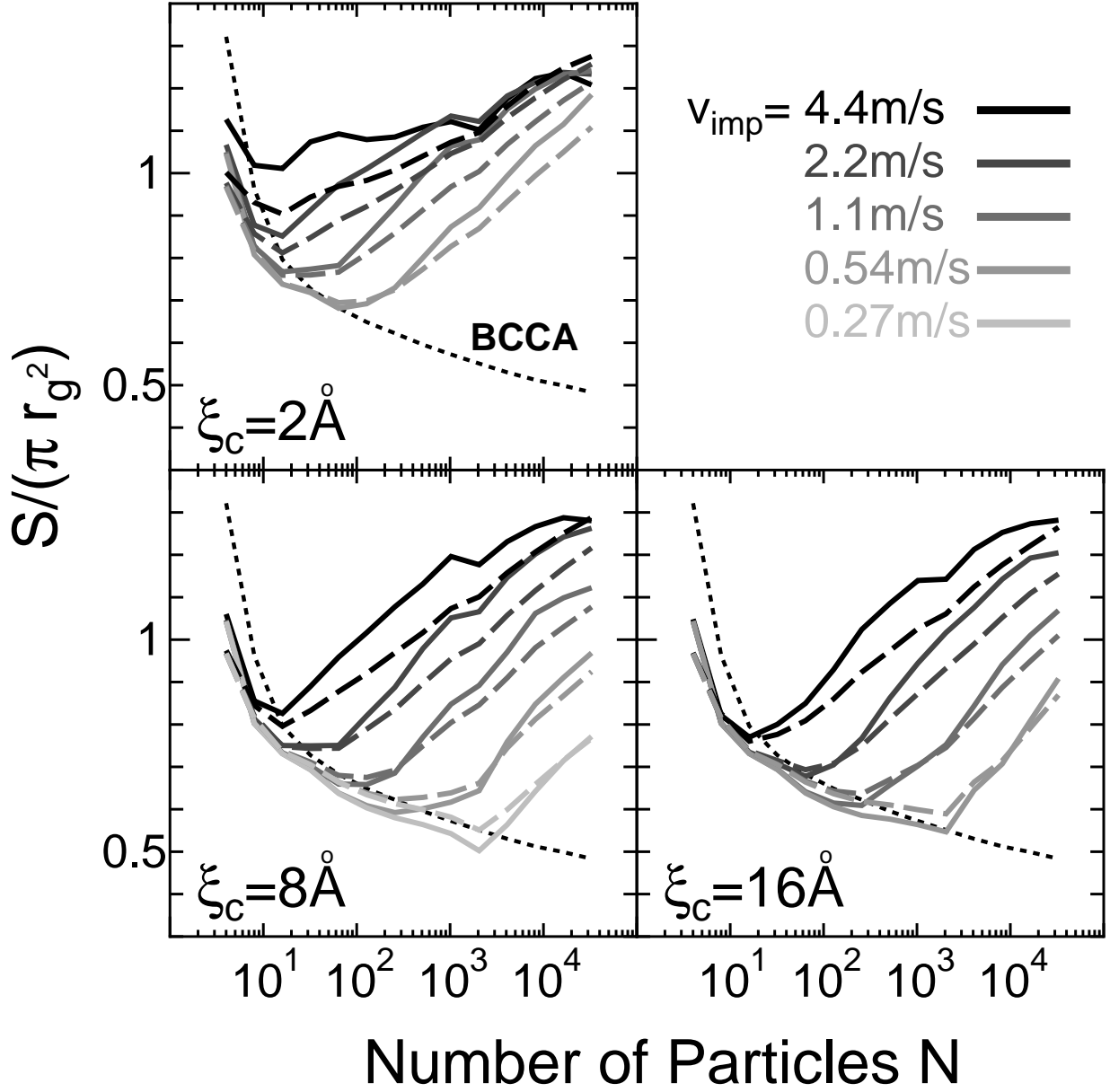}
 \caption{The ratio of $S$ to $\pi r_g^2$ of growing aggregates in the
 simulations for various impact velocities and critical rolling
 displacements. 
The solid lines show the resultant aggregates in our simulations, and
the dotted line indicates that of BCCA clusters.
We also plot the cross sections obtained from Okuzumi et al's expression
(eq.~[\ref{s-O09}]) with dashed lines.
Okuzumi et al's expression reproduces
the cross sections well even for compressed aggregates.
 }
 \label{fig:ssrgheadon}
\end{figure}

\begin{figure}
 \epsscale{1.0}
 \plotone{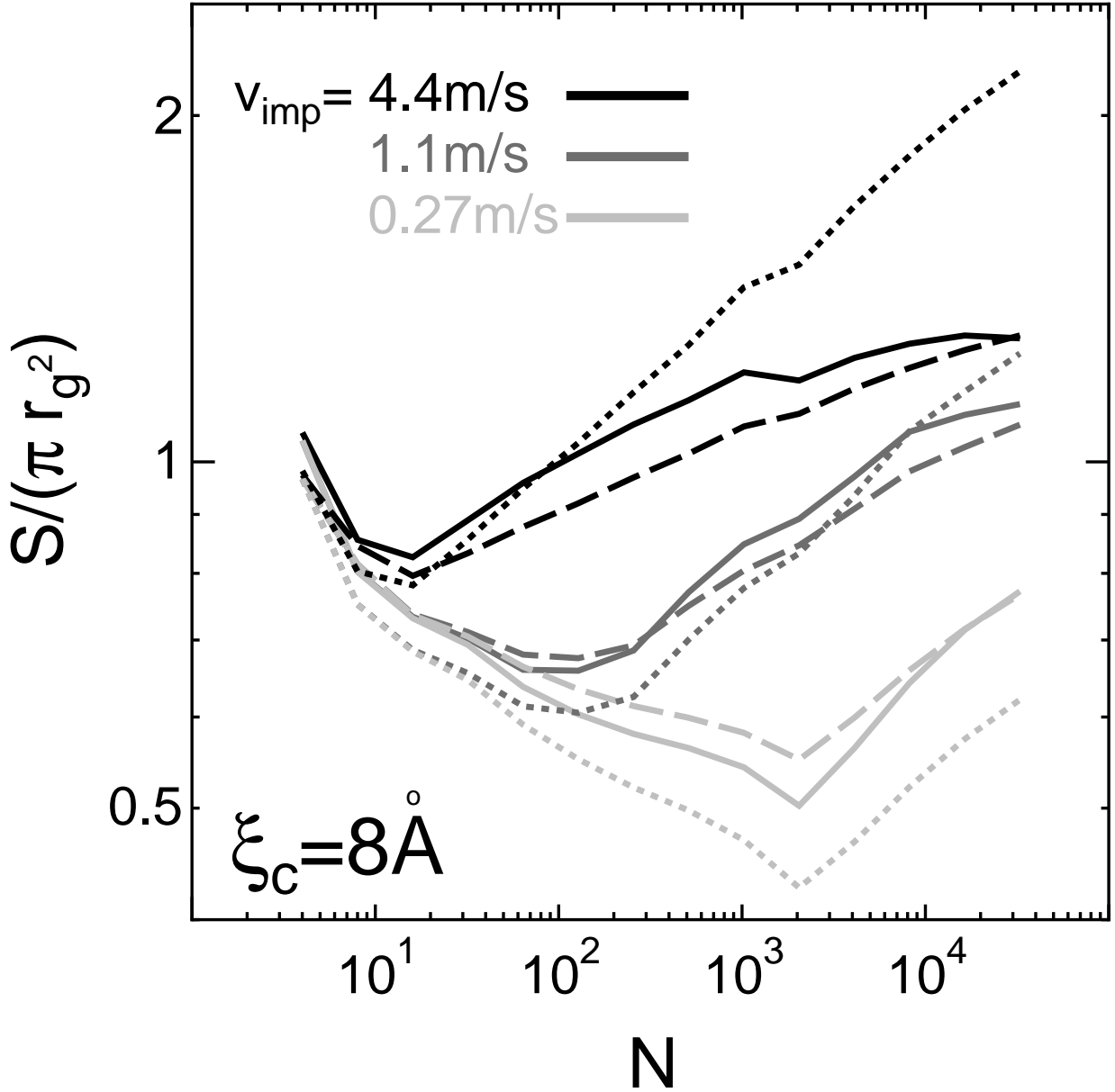}
 \caption{Same as Figure \ref{fig:ssrgheadon} but the prediction
 by the model of Paszun and Dominik (2009) is added.
Only the case of $\xi_\mathrm{c}=8\mathrm{\mbox{\AA}}$ is plotted.
Okuzumi et al's expression reproduces
the cross sections of compressed aggregates better than 
Paszun and Dominik's model.}
 \label{fig:ssrgheadon2}
\end{figure}

In Figure~\ref{fig:ssrgheadon} we plot the ratio of the cross section $S$ 
to $\pi r_g^2$ for all resultant aggregates with solid lines. 
The ratio of  $S$ to $\pi r_g^2$ decreases for small aggregates, 
which is consistent with the BCCA case (dotted lines). 
For sufficiently large aggregates, the ratio increases as a result of
their compression.
%
Okuzumi et al. (2009) proposed a useful expression of 
the cross section $S$ for aggregates formed through hit-and-sticks.
The expression is given by
\begin{equation}
S(r_g, N) = \left(
\frac{1}{S\sub{BCCA}(N)} + \frac{1}{ \pi (5/3) r_g^2}
-\frac{1}{ \pi (5/3) r\sub{$g$, BCCA}(N)^2}
\right)^{-1},
\label{s-O09}
\end{equation}
where the cross section of the BCCA cluster is given by 
(Minato et al. 2006)
\begin{eqnarray}
\frac{S_\mathrm{BCCA}}{\pi r_1^2}=\left\{
 \begin{array}{cc}
  12.5N^{0.685}\exp(-2.53/N^{0.0920}) &(N < 16),\\
  0.352N+0.566N^{0.862} &(N \geq 16).\label{sbcca}
 \end{array}
  \right.
\end{eqnarray}
Equation~(\ref{bccarg}) is used as the expression of
$r\sub{$g$,BCCA}(N)$.
We also plotted the cross sections obtained from equation (\ref{s-O09})
with dashed lines in Figure~\ref{fig:ssrgheadon}.
It is found that the expression by Okuzumi et al. (2009) 
reproduces the cross section surprisingly well for compressed 
aggregates as well as hit-and-stick aggregates, by using the gyration 
radius $r_g$. In this expression, the information of compression
is correctly included through the gyration radius.
Paszun and Dominik (2009) also derived another relation
between $S$ and the aggregate size (i.e., eq.[11] of their paper).
Figure~7 is the same as the left-bottom panel of Figure~6
but the prediction by Paszun and Dominik is also plotted.
Although the prediction by Paszun and Dominik is consistent with
the numerical results,
it overestimates $S$ when the ratio $S/(\pi r_g^2)$ is larger than unity
(i.e., for relatively compact aggregates)
and underestimates for $S/(\pi r_g^2)<0.7$.
The underestimation was also reported by Okuzumi et al.
They found that the underestimation in Paszun and Dominik's model
is severe especially for large and fluffy aggregates.
For other $\xi_\mathrm{c}$,
we also find the same trend as in the case of 
$\xi_\mathrm{c}=8\mathrm{\mbox{\AA}}$ shown in 
Figure~\ref{fig:ssrgheadon2}.
Hence it is concluded that the model of Okuzumi et al.~is 
more accurate than that of Paszun and Dominik.
Once an accurate compression model describing $r_g$ is obtained,
it enables us to calculate the evolution of the cross section with the
help of Okuzumi et al's model.

%
%
\section{COMPRESSION MODEL}
A compression model describing $r_g$ was developed by S08
but it has two limitations.
The model is not directly applicable to low-energy collisions
(i.e., hit-and-stick collisions at the early growth stage) or 
to non-equal-mass collisions.
In order to describe gyration radii and cross sections of aggregates
for both the early hit-and-stick stage and the compression stage seamlessly,
we refine our compression model,
by removing these limitations in a natural way.
Before that, we briefly describe the compression model by S08.

\subsection{Compression Models of W08 and S08}
W08 developed a compression model by introducing the pressure
(or the strength) of aggregates to explain their numerical results on
collisions between BCCA clusters.
The compression model of S08 is based on that of W08.
At a collision of two aggregates with the impact energy 
$E_\mathrm{imp}$, the compression of the merged aggregate from the 
initial volume, $V_\mathrm{initial}$, to the final
volume, $V_\mathrm{final}$, is described in the model of W08 by
\begin{equation}
 E_\mathrm{imp} = -\int_{V_\mathrm{initial}}^{V_\mathrm{final}} P dV.
\label{eimpp}
\end{equation}
The initial volume $V_\mathrm{initial}$ is defined by the volume of the merged
aggregate at the moment that the two aggregates just stick. After the
moment of the sticking, the compression proceeds. The volumes before
and after the compression are evaluated with the radius of gyration,
$r_g$, as in equation~(\ref{vdef}). The pressure $P$ of the aggregates
is given by
\begin{equation}
 P = 2 \left(\frac{5}{3}\right)^6  
\frac{b E_\mathrm{roll}\rho_m}{m_1}
 \left(\frac{\rho}{\rho_m }\right)^{13/3} N^{2/3},\label{pform}
\end{equation}
where the fitting parameter $b$ is set to be 0.15. Note that the
pressure $P$ of the aggregates is dependent on the total number of
constituent particles (or the total mass) as well as the density.
That is, $P$ is not an intensive variable.
This strange property in the pressure comes from the fractal
structure of the aggregates. W08 showed with their simulation of
collisions between BCCA clusters that the compressed aggregates 
have internal structures with a fractal dimension of
2.5. The simulation of sequential collisions done by S08 showed that
their resultant aggregates also have the same fractal dimension of 2.5.

In order to describe the compression of such fractal aggregates,
W08 also introduced the fractal volume defined by
\begin{equation}
 V_f(r_g)\equiv ar_g^{2.5},
\end{equation}
where 
the coefficient $a$ is given by 
$(9\pi/5)^{5/4}\Gamma(9/4) \simeq 7.7$.
Using the fractal volume, the fractal density is defined by
\begin{equation}
 \rho_f(r_g)\equiv\frac{m_1N}{V_f(r_g)}=\frac{m_1N}{a}r_g^{-2.5}
  \label{rhofdef}.
\end{equation}
The dimensions of the fractal volume and the fractal density differ
from those of the ordinary volume and density. These fractal
quantities are related with the ordinary quantities $V$ and $\rho$ as
\begin{eqnarray}
\frac{V(r_g)}{v_m}&=&\left(\frac{5}{3}\right)^{3/2}
 \left( \frac{V_f(r_g)}{v_{f,1}}\right)^{6/5},\label{ord_fra_v}\\
 \frac{\rho(r_g)}{\rho_m}&=&\left(\frac{3}{5}\right)^{3/2} 
  \left(\frac{\rho_f(r_g)}{\rho_{f,1}}\right)^{6/5}
 N^{-1/5},\label{ord_fra}
\end{eqnarray}
where $v_m$ ($=4/3\pi r_1^3$) is the volume of a monomer and $v_{f,1}$ is
given by $ar_1^{2.5}$.
Using the fractal volume $V_f$ (and the fractal density $\rho_f$),
equation~(\ref{eimpp}) is rewritten as
\begin{equation}
 E_\mathrm{imp} = -\int_{V_{f,\mathrm{initial}}}^{V_{f,\mathrm{final}}} P_f
  dV_f,
  \label{eimppf}
\end{equation}
where the fractal pressure $P_f$ is given by
\begin{equation}
 P_f \equiv P \frac{dV}{dV_f}
     =  4 \frac{bE_\mathrm{roll}\rho_{f,1}}{m_1}
     \left(\frac{\rho_f}{\rho_{f,1}}\right)^5.\label{pf} 
\end{equation}
It should be noticed that the fractal pressure is dependent on
$\rho_f$ but not on the total mass. 
That is, $P_{f}$ is an intensive variable.
Equation (\ref{eimppf}) (or
[\ref{eimpp}]) reproduces the numerical results on the compression
at collisions between BCCAs.

S08 pointed out that W08's compression model
needs a minor modification to describe the compression of
partially compressed aggregates at their collisions, which occurs
in their simulation of sequential collisions. At the moment of
sticking at each collision, large voids are produced in the merged
aggregate. The volume of the new voids is included in the initial
volume of the merged aggregate, $V_{f,\mathrm{initial}}$ in
equation~(\ref{eimppf}).
The energy required for compression of the new voids is
$\sim E_\mathrm{roll}$ and it is much smaller that that predicted by
equation~(\ref{eimppf})
at collisions between partially compressed aggregates. To describe
the compression at such collisions, S08 modified W08's
model. Since the energy required for the crush of the new voids is
negligible, the initial fractal volume of the merged aggregate in
equation~(\ref{eimppf}) is set to be the sum of the fractal volumes of two
colliding aggregates, by removing the volume of the new voids.
That is,
\begin{equation}
 V_{f,\mathrm{initial}} = V_{f,1} + V_{f,2}\label{vfinit}. 
\end{equation}
Using equations~(\ref{eimppf})-(\ref{vfinit}), we have the (final)
fractal density of the merged aggregate, $\rho_{f,\mathrm{final}}$,
produced at collisions of two equal-mass aggregates with the fractal
density $\rho_{f,0}~(=Nm_1/[V_{f,1}+V_{f,2}])$
\begin{equation}
 \left(\frac{\rho_{f,\mathrm {final}}}{\rho_{f,1}}\right)^4=
  \left(\frac{\rho_{f,0}}{\rho_{f,1}}\right)^4
   +\frac{E_\mathrm{imp}}{bNE_\mathrm{roll}}.\label{oldmodel}
\end{equation}
where $N$ is the number of constituent particles in the merged one.
Equation~(\ref{oldmodel}) describes the density evolution of partially
compressed aggregates growing through mutual collisions.
Equation~(\ref{oldmodel}) with $b=0.15$ reproduces 
the density evolution of growing aggregates in Figure~\ref{fig:rgrho}
for $N>N_\mathrm{crit}$, as seen in Figure~8 of S08.
\subsection{Refinement of the Compression Model}

The compression model by S08 is not applicable to the early growth stage 
($N<N_\mathrm{crit}$) where the compression is ineffective. 
This is because the energy
required for the crush of the new voids is neglected in the model.
This limitation is removed, by taking into account the compression process 
of the new voids produced at sticking of two aggregates.
Furthermore, the model of S08 assumes the collisions of equal-mass aggregates.
We also remove this limitation in a reasonable way.
At the extension of our compression model to non-equal-mass collisions, 
we assume that the compressed aggregates have the fractal dimensions 
of 2.5 as well as in the case of equal-mass collisions. 
The validity of the assumption will be discussed in Section 5.

\begin{figure}
 \plotone{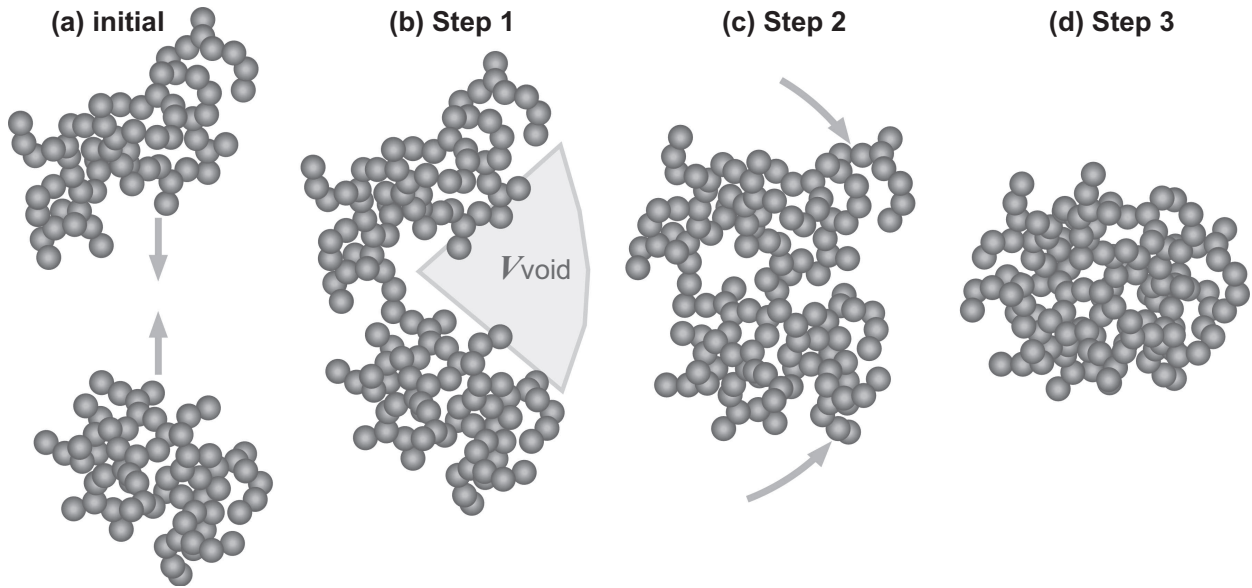} 
 \caption{
Schematic illustration of three steps in the refined compression model. 
(a)Before a collision between two aggregates.
(b)Step~1: At the moment of their stick. New voids are created in the 
merged aggregate.
(c)Step~2: Compression of the new voids.
(d)Step~3: Further compression of the merged aggregate.
 }
 \label{fig:refinedmodel}
\end{figure}

Density evolution of aggregates at each collisions
is divided into the following three steps: 
\begin{enumerate}
 \item Creation of new voids at the sticking of two aggregates.
 \item Compression of the new voids in the merged aggregate. 
 \item Further compression of the merged aggregate after the crush of
       the new voids.
\end{enumerate}
These steps are schematically explained by Figure~\ref{fig:refinedmodel}.
We describe the change in the volume or the 
density of the aggregates at each step in detail below.

In Step~1, the density decreases because of the new voids in the merged
aggregate. The change in the density
is described by Okuzumi et al. (2009).
When the two aggregates with the masses $M_1$, $M_2$, and the
volumes $V_1$, $V_2$ ($\le V_1$) collide with each other, the volumes
of the merged aggregate $V'_{1+2}$ (before the compression)
is given by
\begin{equation}
 V'_{1+2} = V_1 + V_2 + V_\mathrm{void},\label{vdash}
\end{equation}
where the volume of the voids $V_\mathrm{void}$ is obtained from
the empirical formula
\begin{equation}
 V_\mathrm{void} =\min\left[
                       \chi\sub{BCCA}-1.03\ln\left(\frac{V_1+V_2}{2V_2}\right),
                       6.94\right]V_2
\end{equation}
(see Okuzumi et al. [2009] for the derivation).
The collision of equal-mass
aggregates is not assumed in Okuzumi et al. (2009).
Note that $\chi\sub{BCCA} (\equiv 2^{3/d_f} -2)$ is 0.83 
for BCCA clusters formed through head-on collisions having the
fractal dimension $d_f=2.0$.
The density of the merged aggregates before the compression
is given by ($M_1+M_2)/V'_{1+2}$. The fractal volume $V'_{f,1+2}$
and the fractal density $\rho'_{f,1+2}$ are obtained from
equations~(\ref{ord_fra_v}) and (\ref{ord_fra}), respectively.


In Step~2 and 3, the merged aggregate is compressed
and the density increases. Step~2 is the compression of the
new voids. We put the energy required for the crush of the
new voids to be $b'E_\mathrm{roll}$, where $b'$ is the non-dimensional
parameter. By fitting with results of the simulation,
this parameter is fixed to be $b'=3b$, as will be shown in the next
subsection. This relation would be reasonable because both parameters 
are related to the beginning of compression.

When the impact energy $E_\mathrm{imp}$ is smaller than
$b'E_\mathrm{roll}$,
the new voids are only partially compressed at Step~2.
In this case, the compression at Step 3 does not occur 
because of the small impact energy.
The impact energy $E_\mathrm{imp}$ is 
evaluated using the reduced mass, $(1/M_1+1/M_2)^{-1}$.
In this case, the decrease in the fractal volume of the merged
aggregates, $\Delta V_{f,\mathrm{1+2}}$, is evaluated to be
\begin{equation}
 \Delta V_{f,1+2} = \frac{E_\mathrm{imp}}{b'E_\mathrm{roll}}
  V_{f,\mathrm{void}},
\end{equation}
where
\begin{equation}
 V_{f,\mathrm{void}} = V'_{f, 1+2} - V_{f,1} - V_{f,2}.
\label{vfvoid}
\end{equation}
It should be noticed that the fractal volume $V_{f,\mathrm{void}}$
defined by equation~(\ref{vfvoid}) is negative in collisions where
the volume ratio $V_1/V_2$ is larger than $8 \times 10^4$.
A prescription for such high-volume-ratio collisions will be described 
at the end of this subsection.
Here we consider the case where $V_{f,\mathrm{void}}$ is positive.
Then the final fractal volume $V_{f, 1+2}$ after the compression
is given by
\begin{equation}
 V_{f, 1+2} = V'_{f, 1+2} - \Delta V_{f,1+2}.\label{modellow}
\end{equation}
and the final fractal density $\rho_{f,\mathrm{final}}$ of the merged
aggregate is given by $(M_1+M_2)/V_{f, 1+2}$.

When the impact energy is larger than $b'E_\mathrm{roll}$, 
the new voids are crushed completely and the merged
aggregate is further compressed at Step~3. Since the fractal
volume of the merged aggregate is $V_{f,1}+V_{f,2}$ at the end
of Step~2, the final fractal volume $V_{f, 1+2}$ after Step 3
is obtained from the equation
\begin{equation}
E_\mathrm{imp} - b'E_\mathrm{roll} =
 - \int_{ V_{f,1}+V_{f,2} }^{ V_{f, 1+2} } P_f(\rho_f) dV_f.
\label{modelhigh}
\end{equation}
Integrating the RHS of equation~(\ref{modelhigh}), we obtain the final
fractal density for $E_\mathrm{imp} > b'E_\mathrm{roll}$ as
\begin{equation}
 \left(\frac{\rho_{f,\mathrm {final}}}{\rho_{f,1}}\right)^4=
  \left(\frac{\rho_{f,0}}{\rho_{f,1}}\right)^4
   +\frac{E_\mathrm{imp}-b'E_\mathrm{roll}}{bNE_\mathrm{roll}},
   \label{rhoevol_up1}
\end{equation}
where we used 
\begin{equation}
\rho_{f,0}= \frac{M_1+M_2}{V_{f,1}+V_{f,2}}.
\label{rhof0}
\end{equation}
In the limit of
$E_\mathrm{imp}\gg b'E_\mathrm{roll}$,
equation~(\ref{rhoevol_up1}) is identical to
equation~(\ref{oldmodel}) (or the compression model by S08).
The final fractal volume $V_{f,1+2}$ is given by
$(M_1+M_2)/\rho_{f,\mathrm{final}}$.
>From the fractal density, we obtain the gyration radius $r_g$,
using equation~(\ref{rhofdef}). The cross section $S$
is also obtained from Okuzumi et al's expression (eq.[\ref{s-O09}]).
In this way, we can calculate the density evolution 
(i.e., the evolution of $r_g$ and $S$) at both low- and high-energy 
collisions, using equation~(\ref{modellow}) for 
$E_\mathrm{imp} < b'E_\mathrm{roll}$ and equation~(\ref{rhoevol_up1}) for 
$E_\mathrm{imp} > b'E_\mathrm{roll}$.

In high-volume-ratio collisions where $V_1/V_2 > 8\times10^4$,
as noticed above, $V_{f,\mathrm{void}}$ is negative and
a special prescription is necessary. 
Since a negative $V_{f,\mathrm{void}}$ means no voids,
Step 2 should be omitted and 
the merged aggregate is compressed only with Step 3.
That is, equations~(\ref{modelhigh})-(\ref{rhof0}) are used
for all impact energies in this case.
In equations~ (\ref{modelhigh})-(\ref{rhof0}), the terms of 
$b'E_\mathrm{roll}$ is omitted and $V_{f,1}+V_{f,2}$ is replaced by 
$V'_{f,1+2}$ because Step 2 does not occur.

\subsection{Test of the Refined Compression Model}

\begin{figure}
 \plotone{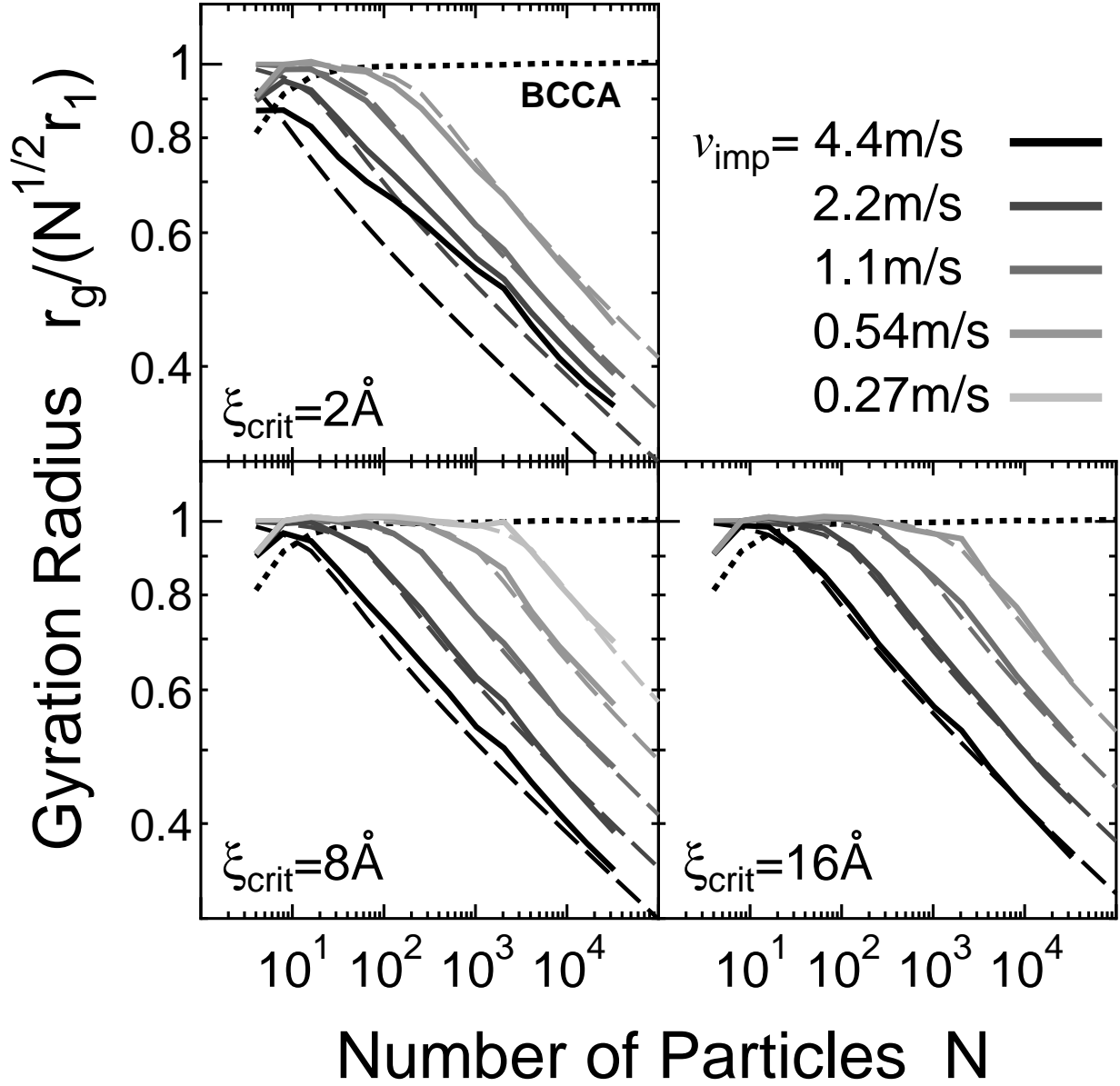}
 \caption{Evolution of the gyration radius $r_g$ 
calculated with the refined compression model.
The model curves are plotted with dashed lines.
The parameters are set as $b = 0.15$ and $b'=3b$.
 The model curves agree well with 
 the numerical results (solid lines) for all size range
in all cases.
}
 \label{fig:rgfit}
\end{figure}

\begin{figure}
 \plotone{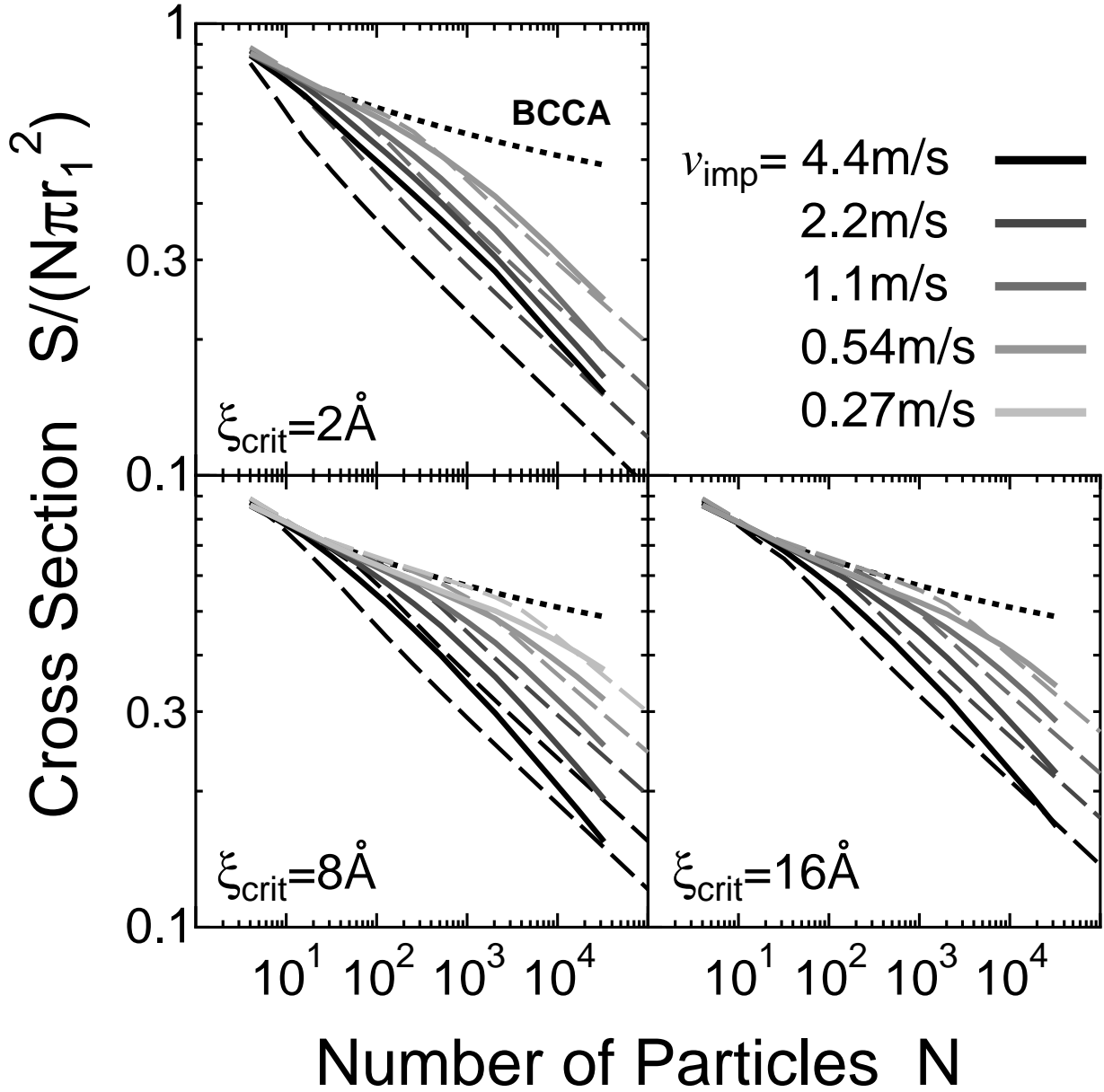}
 \caption{Evolution model of the cross section $S$ 
obtained from equation~(\ref{s-O09}) and
the gyration radius calculated with the refined model 
(dashed lines).
The refined compression model also reproduces the cross sections
in the numerical simulations (solid lines) with the help of
equation~(\ref{s-O09}).
}
 \label{fig:sfitoku}
\end{figure}

Let us test the refined compression model with the numerical results.
Using the refined compression model, we calculate the evolution of
gyration radius for the same condition as the numerical simulations by
S08 and also obtain the cross sections of the aggregates with 
equation~(\ref{s-O09}). The results are shown in Figures~\ref{fig:rgfit}
and \ref{fig:sfitoku}. In Figure~\ref{fig:rgfit}, we plot the evolution
of the gyration radius calculated with the refined compression model
and compared it with the numerical results by S08. The parameters $b$ and 
$b'$ are set to be $b=0.15$ and $b'=3b~(=0.45)$, respectively.
With this setting of the parameters, the refined model reproduces well 
the numerical results at both the early growth stage and the compression 
stage. Figure~\ref{fig:sfitoku} shows the evolution of the cross sections
and indicates that the refined model also succeeds in describing the 
cross sections with the help of equation~(\ref{s-O09}). 

\begin{figure}
 \plotone{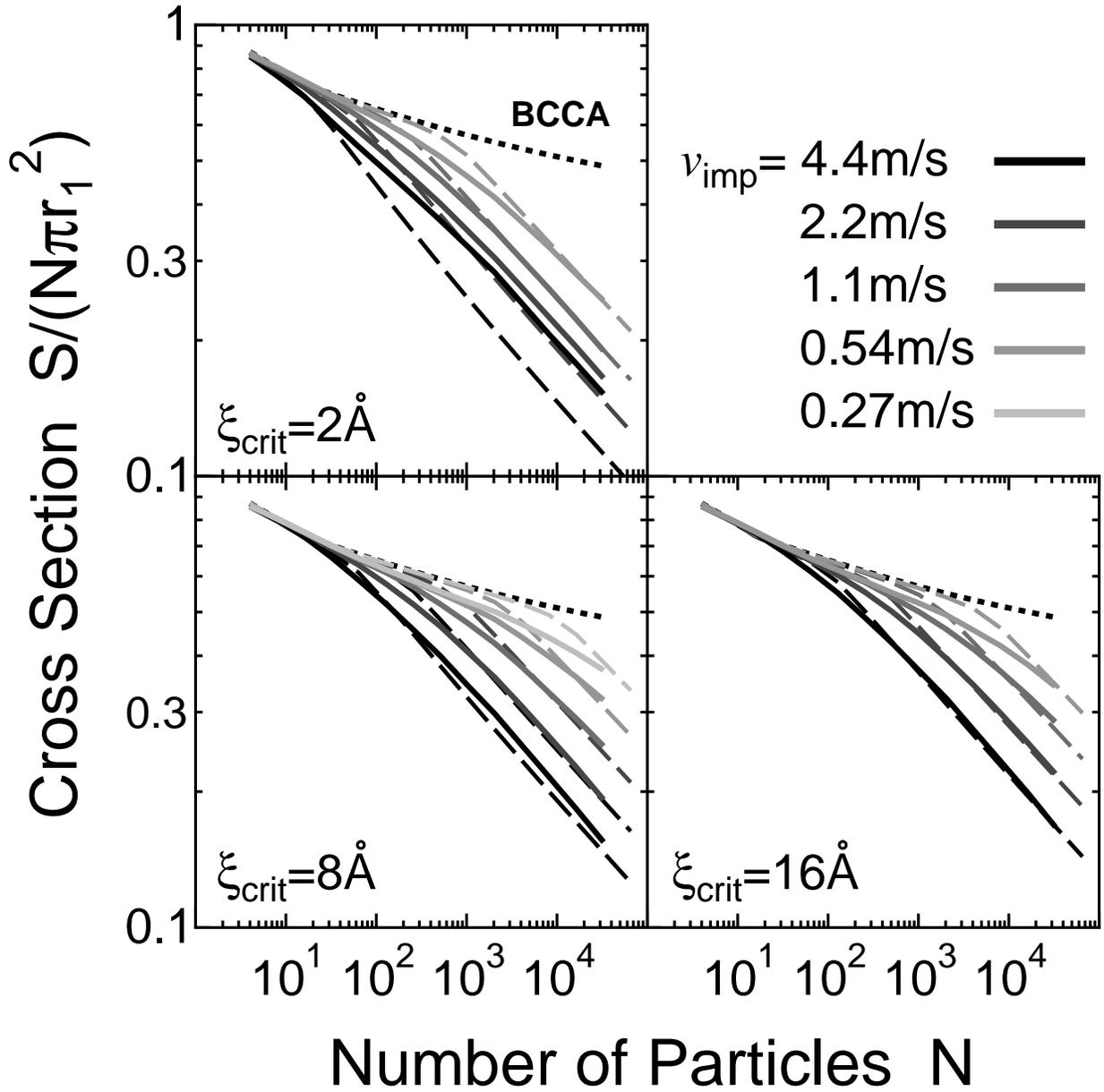} 
 \caption{Same as Fig.~\ref{fig:sfitoku}
 but the model cross sections are calculated with the refined model
 in the direct way, using the characteristic size $r_S$.
 With the direct way, we can describe accurately the evolution of the 
 cross section with the parameter $b = 0.60$.
}
 \label{fig:sfit}
\end{figure}

In the above, the evolution of gyration radius of growing aggregates
are calculated with the refined compression model
and their cross sections are indirectly calculated, by using
$r_g$ and equation~(\ref{s-O09}).
We also propose another way to describe the cross sections of aggregates.
We define alternative characteristic sizes of aggregates $r_S$ by 
\begin{equation}
r_S=\sqrt{S/\pi}.
\label{eq:rs}
\end{equation}
It would be possible to describe the evolution of $r_S$
directly (instead of the gyration radius)
with the refined compression model in the following way.
Using this characteristic size $r_S$ instead of $\sqrt{5/3}r_g$,
we can define the volume and the bulk density of the aggregate
by the similar equations to (\ref{vdef}) and (\ref{rhodef}).
The fractal volume and the fractal density are also defined in the same way.
Then, applying the refined compression model to the fractal density
defined with $r_S$, we can describe the evolution of $r_S$
as well as in the case of $r_g$. 
The evolution of the cross section $S$ is calculated with 
equation~(\ref{eq:rs}).
This is a direct way to describe the
cross section rather than the above.
In this calculation of $S$,
we have to be cautious with the following two points.
One is the modification in equation~(\ref{vdash}).
At a collision of sufficiently fluffy aggregates, 
equation~(\ref{vdash}) can give the volume $V'_{1+2}$ larger
than that of the BCCA cluster with the same mass, $V_\mathrm{BCCA}$
when the size $r_S$ is used instead of $r_g$.
Such a large $V'_{1+2}$ is not realistic. In this case, we set
the volume $V'_{1+2} = V_\mathrm{BCCA}$ instead of equation~(\ref{vdash}).
The other point is the parameter $b$.
Although $b$ is set to be 0.15 in the case of $r_g$,
we have to calibrate the parameter $b$ again in the case of $r_S$
as a result of the fitting with numerical results.
In Figure~\ref{fig:sfit}, we plot the evolution
of $S$ calculated with this direct way.
In this calculation, the parameters are set to be $b=0.6$ and $b'=3b$.
We see that the refined model works well for the evolution of the
cross sections with this direct way, too.\footnote{
In the early growth stage of Figure~\ref{fig:sfit},
we used $V'_{1+2} = V_\mathrm{BCCA}$
instead of equation~(\ref{vdash}) 
when the volume of equation~(\ref{vdash})
is larger than $V_\mathrm{BCCA}$, as mentioned above.
}

It is found that the refined compression model enables us to describe 
the whole evolution of the radius of gyration $r_g$ and the cross 
section $S$ of growing aggregates.
Note that the refined model is applicable to non-equal-mass
collisions though the tests in Figure~\ref{fig:rgfit}-\ref{fig:sfit} are 
done only in the equal-mass case. 
At the extension of the refined model to non-equal-mass
collisions, the fractal dimension of compressed aggregates
is assumed to be 2.5 even in the case of non-equal-mass collisions
though it is not verified with $N$-body simulations in non-equal-mass cases.
It is possible that a very large mass ratio increases the fractal dimension,
as seen in BPCA clusters.
Okuzumi et al. (2009) examined the effect of the mass ratio
on the fractal dimension for aggregates growing with hit-and-stick
collisions with $N$-body simulations. They showed that
the collisions with the mass ratio of 10 increases
the fractal dimension $d_f$ only by 0.1 (see their figure 6). Furthermore, 
collisions with such a mass ratio have a major contribution at dust 
growth in protoplanetary disks, as shown by Okuzumi et al.~(2009,2012).
Hence the effect of non-equal-mass collisions would not 
change largely the fractal dimension of compressed aggregates
in the realistic growth process.

\begin{figure}
 \plotone{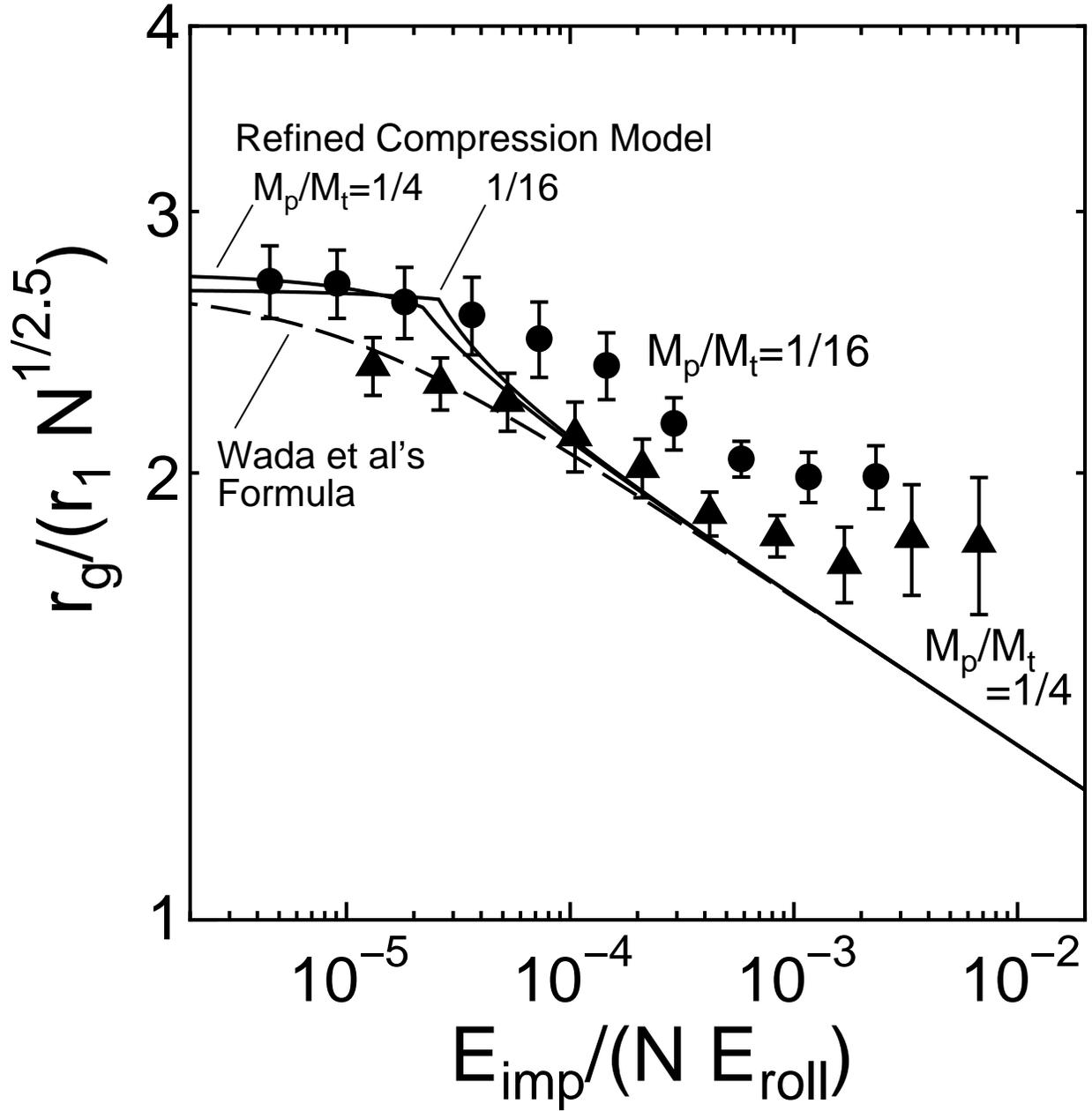}
\caption{Test by compression in non-equal-mass aggregate collisions.
The results of $N$-body simulations of non-equal-mass BCCA collisions
are plotted by circles and triangles for
the mass ratios $M_p/M_t =1/16$ and $1/4$, respectively.
Each data point is the averaged value obtained from 10 runs of $N$-body 
simulation for different types of initial BCCA clusters, as done by W08.
The standard errors in the averaged values are also displayed.
The predictions by the refined model are plotted by solid 
lines and the dashed line indicates the formula obtained by W08.
}
 \label{fig:additional}
\end{figure}

To confirm the validity of the refined model in non-equal-mass
collisions, we have further performed additional $N$-body simulations of 
aggregate collisions. Similar to W08, we consider collisions of two BCCA 
clusters but their masses are not equal in the present case.
The projectile BCCA cluster consists of 1024 particles (or 4096 particles)
while the number of constituent particles of the target BCCA cluster 
is 16384. Their mass ratio is $1/16$ (or $1/4$).
The constituent particles are icy ones with the radius with 0.1$\mu$ m.
The impact velocity $v\sub{imp}$ is a parameter.
We set $v\sub{imp} \le 4.4$ms$^{-1}$ since we focus on the compression 
process rather than fragmentation (Wada et al. 2007,2008).
The numerical results of compression at the non-equal-mass collisions
are shown in Figure~\ref{fig:additional}.
The predictions by the refined model with $b=0.15$ are plotted by solid 
lines and the dashed line indicates the formula by W08 (their equation~[45]).
The numerical results in the non-equal-mass collisions approximately 
agree with the predictions by the refined model though upper shifts by 
$\sim$20\% are observed in the case of $M_p/M_t=1/16$.
As well as in the case of equal-mass collisions,
the slope in $r_g$-$E\sub{imp}$ relation is approximately given by -0.1,
which indicates that compressed aggregates have the fractal dimension of 
2.5. The upper shifts in the numerical results indicate that the 
compression requires larger impact energy in non-equal-mass collisions 
than the equal-mass case. This effect in non-equal-mass collisions would 
be included by adopting a larger parameter $b$ in the refined model.
Furthermore, for high-mass-ratio collisions with $M_t/M_p \gg 10$,
the refined model is not verified through $N$-body simulations yet
although such collisions have only a minor contribution in dust 
growth (Okuzumi et al.~2009,2012).
In the future work, the effect of non-equal-mass collisions
should be further examined in the numerical simulation of sequential 
collisions as done by S08 in order to calibrate the parameter
$b$ more accurately.
\section{SUMMARY}
We examined the evolution of the geometrical cross section of the
growing (icy) aggregates obtained by $N$-body simulations of
sequential head-on collisions (S08)
and constructed to construct a refined compression model,
which is applicable to the description of the evolution of both
geometrical cross sections and gyration radii of growing aggregates.
The results are summarized as follows:
\begin{enumerate}
 \item We examined geometrical cross sections of the aggregates
       produced in the simulation of sequential collisions done in our
       previous paper. 
	As aggregates grow, compression becomes effective and makes their 
	cross sections smaller than those of the BCCA clusters. The 
	beginning of the compression is given by equation~(\ref{ncrit}), 
	as seen in the evolution of the gyration radius.
 \item The relation between the cross section and the gyration radius
       seen in aggregates obtained by S08
	is well described by Okuzumi et al's expression.
       This indicates that Okuzumi et al's expression is valid
        for compressed aggregates as well as hit-and-stick aggregates.
       If the evolution of the gyration radius is well described by
       a compression model, Okuzumi et al's expression enables us to
       calculate the cross section, too.
 \item We further refined the compression model of S08, by including 
       the compression energy for the voids produced at the sticking 
       of two aggregates. The refined model is also extended to
       non-equal-mass collisions in a reasonable way.
	With the refined model, we can accurately reproduce the evolution 
        of both the gyration radius and the cross section of aggregates
        obtained by S08 from their early growth stage.
       The validity of the refined compression model
       for non-equal-mass collisions is also checked by
       additional numerical simulations of BCCA collisions.
       Although S08 considered only icy aggregates in the numerical 
       simulation, our compression model would be also applicable to 
       silicate aggregates by using a suitable value of $E_\mathrm{roll}$. 
\end{enumerate}

Our $N$-body simulations of aggregate collisions and the refined 
compression model indicate that collisional compression is not so 
effective. As a result, dust aggregates (or initial planetesimal material) 
would have extremely low bulk densities, as suggested by S08. Okuzumi 
et al. (2012) showed that such extremely low bulk densities of aggregates 
accelerate their growth in protoplanetary disks and help their overcoming 
of the radial drift barrier against the planetesimal formation. However, 
solar-system bodies do not have such low densities at present. Dust 
aggregates (or planetesimals) should be compressed by other processes. 
A steady ram pressure due to the gas drag on aggregates and a 
self-gravity of sufficiently large aggregates would be candidates for 
aggregate compression, as indicated by S08 and Okuzumi et al. (2012). 
For relatively compact dust cakes 
($\rho \sim0.1$g$/$cm$^3$) made of micron-sized silicate particles, 
compression is observed at a pressure $> 100$Pa (Blum and 
Schr\"{a}pler 2004). However, for icy aggregates with very low bulk 
densities ($\rho \ll 0.1$g$/$cm$^3$), compressive strength 
has not yet been measured. In future work, compression strength of 
very fluffy aggregates should be measured in numerical simulations 
and laboratory experiments.

In the present study,
we focus on the aggregates obtained at head-on collisions. 
At oblique collisions, the merged aggregates are elongated
(W07; Paszun and Dominik 2009). Although our model
indicates inefficient compression at aggregate collisions, the effect of 
oblique collisions would further hinder compression.
In future work, we should clarify the validity of our compression model 
in the case where oblique collisions are included.
\acknowledgments
The authors would like to thank Hiroshi Kimura, 
Tetsuo Yamamoto and Hiroshi Kobayashi for
their valuable comments. We would also like to thank Takeshi Chigai for
technical support with respect to the computer setup. This study was
supported by a Grant-in-Aid from JSPS(22540242, 22740299).


\end{document}